\documentclass[aps,prx,reprint,superscriptaddress,floatfix, longbibliography]{revtex4-2}

\usepackage[english]{babel}
\usepackage{graphicx}
\usepackage{amsmath}
\usepackage{amssymb}
\usepackage{bm}
\usepackage{xcolor}
\usepackage{booktabs}

\begin{document}

\title{Logarithmic Quasi-Long-Range Order and Novel Continuous Phase Transition \\ in the Two-Dimensional XY Model with $1/r^4$ Interaction}

\author{Xianzhi Pan}
\affiliation{Hefei National Research Center for Physical Sciences at the Microscale and School of Physical Sciences, University of Science and Technology of China, Hefei 230026, China}
\author{Zhijie Fan}
\email{zfanac@ustc.edu.cn}
\affiliation{Hefei National Research Center for Physical Sciences at the Microscale and School of Physical Sciences, University of Science and Technology of China, Hefei 230026, China}
\affiliation{Hefei National Laboratory, University of Science and Technology of China, Hefei 230088, China}
\affiliation{Shanghai Research Center for Quantum Science and CAS Center for Excellence in Quantum Information and Quantum Physics, University of Science and Technology of China, Shanghai 201315, China}
\author{Youjin Deng}
\email{yjdeng@ustc.edu.cn}
\affiliation{Hefei National Research Center for Physical Sciences at the Microscale and School of Physical Sciences, University of Science and Technology of China, Hefei 230026, China}
\affiliation{Hefei National Laboratory, University of Science and Technology of China, Hefei 230088, China}
\affiliation{College of Physics, Guizhou University, Guiyang 550025, China}

\date{\today}

\begin{abstract}
We investigate the phase diagram and critical properties of the two-dimensional classical XY model with interactions decaying as \(1/r^{2+\sigma}\).
At the marginal case \(\sigma=2\), we show that the low-\(T\) phase is characterized by logarithmic quasi-long-range order (log-QLRO), where spin correlations decay algebraically with the logarithm of distance as \(C(r)\sim(\ln r)^{-\eta_\ell}\) with \(\eta_\ell\propto T\), and that the transition into this phase is a continuous transition beyond both the Ginzburg-Landau-Wilson paradigm and the Berezinskii-Kosterlitz-Thouless mechanism.
Our analysis combines a near-exact Gaussian spin-wave theory, an adiabatic renormalization-group analysis, and large-scale Monte Carlo simulations.
We find that the \(1/r^4\) interaction logarithmically modifies the spin-wave stiffness kernel as \(\gamma(k)\simeq\kappa k^2\ln(1/k)\), 
which suppresses vortex proliferation and gives rise to a nonuniversal correlation-length exponent \(\nu \simeq 1/\sqrt{2\pi\kappa}\).
Crucially, our simulations reveal that the short-range algebraic QLRO phase is unstable against an arbitrarily weak long-range perturbation for \(\sigma\le2\).
These results establish \(\sigma=2\) as the crossover boundary between the long-range and short-range universality classes and resolve the long-standing controversy over the phase diagram.
\end{abstract}
\maketitle
\section{Introduction}
\label{sec:intro}
The two-dimensional (2D) XY model has long served as a paradigmatic example of a topological phase transition. 
The Mermin-Wagner theorem forbids spontaneous symmetry breaking at any finite temperature ($T$), yet the system undergoes a Berezinskii-Kosterlitz-Thouless (BKT) transition driven by the unbinding of vortex-antivortex pairs~\cite{Berezinskii1971,Kosterlitz1973}. The low-temperature (low-\(T\)) phase is characterized by algebraic quasi-long-range order (QLRO), where the spin correlation decays 
algebraically with distance as $C(r)\sim r^{-\eta}$, with exponent $\eta \propto T$. 
Above the critical temperature $T_c$, free vortices proliferate, and the system enters a disordered phase.
The BKT transition represents one of the most experimentally verified predictions in condensed matter and statistical physics. Its signatures have been observed in physical systems including superconducting films~\cite{Verma2024,Mondal2011}, superfluid $^4$He films~\cite{Bishop1978},
ultracold atomic gases~\cite{Hadzibabic2006,Fletcher2015,Sunami2022,Sunami2023}, Josephson junction arrays~\cite{Resnick1981}, and layered superconductors~\cite{Baity2016}. The role of vortex-core energy in layered superconductors has also been analyzed theoretically~\cite{Benfatto2007}.

Long-range (LR) systems with algebraically decaying interactions \(1/r^{2+\sigma}\) have attracted growing interest due to advances in quantum simulators, which offer precise control over interaction range and strength. LR Ising and XY models have been realized in trapped ions~\cite{Monroe2021,Britton2012} and 2D Rydberg arrays~\cite{Chen2023}, complemented by laser-painted cavity-mediated interactions in quantum gases~\cite{Bonifacio2024} and dipolar interactions in ultracold polar molecules~\cite{Yan2013} and magnetic atoms~\cite{Lu2012,Aikawa2012}. 
These experimental capabilities raise the question of how LR interactions reshape the BKT transition and the low-\(T\) QLRO phase.

Despite decades of theoretical and numerical effort, the phase diagram of the 2D LR XY model remains debated, with three main scenarios. The first scenario, proposed by Giachetti \textit{et al.}~\cite{Giachetti2021} and reviewed in Ref.~\cite{Defenu2023} on the basis of perturbative functional renormalization analysis combined with Sak's criterion~\cite{Fisher1972,Sak1973}, posits that for \(7/4 < \sigma < 2\) the system enters an intermediate algebraic QLRO phase through a BKT transition and then, upon further cooling, develops true LRO through a symmetry-breaking transition at a lower temperature.
Another scenario, due to Walther \textit{et al.}~\cite{Walther2026}, argues that the BKT transition persists for all \(\sigma\), even as $\sigma \to 0$: spin waves renormalize the vortex-antivortex interaction in the presence of LR couplings, and Landau-Peierls-type arguments predict an intermediate QLRO phase followed by a low-\(T\) LRO phase.
A third scenario, supported by large-scale Monte Carlo simulations~\cite{Xiao2025LRXY,Yao2025LRXY},
locates the crossover at $\sigma_* = 2$: for $1 < \sigma < 2$, the system undergoes a single second-order transition into a ferromagnetic LRO phase with continuously varying critical exponents, whereas for $\sigma > 2$ standard BKT physics is recovered. The low-\(T\) LRO phase for $\sigma < 2$ exhibits Goldstone-mode fluctuations, giving $C(r) \sim C_0 + a r^{-(2-\sigma)}$.

Recent advances in the field-theoretic treatment have provided crucial insights into this controversy. The $\epsilon$-expansion for LR $O(N)$ models~\cite{Li2026Ising} in dimension $(4-\epsilon)$ reveals that for $\sigma < 2$, the LR interaction is relevant: the SR Wilson-Fisher fixed point (SR-WFP) becomes unstable, and a nontrivial LR Wilson-Fisher fixed point (LR-WFP) emerges with the anomalous dimension
\begin{align}
\eta = 2-\sigma + \frac{N+2}{(N+8)^2}\frac{(\epsilon - 2\delta)^3}{2\epsilon - 3\delta} + O(\epsilon^3), 
\label{eq:eta_expansion}
\end{align}
where $0<\delta=2-\sigma <\epsilon/2$.
It is implied that the deviation from the LR Gaussian value $2-\sigma$ can exist, 
and, for $\sigma\to 2$, this deviation does not vanish but smoothly reduces to the celebrated SR Wilson-Fisher value. 
Historically, Sak assumed that $\eta$ remains locked at $2-\sigma$,
and proposed the $\sigma_*=2-\eta_{\rm SR}$ boundary to resolve the discontinuity at $\sigma=2$.
Equation~\eqref{eq:eta_expansion} suggests a possibility that neither the assumption underlying Sak's criterion
nor the question it was to address exists.
Li \textit{et al.} extended this analysis to LR quantum $O(N)$ models near $d=3$~\cite{Li2026Quantum} and to LR percolation and Lee-Yang models near $d=6$~\cite{Li2026Percolation}, consistently finding a nontrivial LR-WFP for $\sigma<2$. Although the $\epsilon$-expansion at $d=2$ is not quantitatively reliable, it provides a qualitative framework for the phase diagram.
High-precision numerical studies of various LR systems, including Ising, percolation, XY, Heisenberg and loop-erased random walks~\cite{Liu2025,Xiao2025LRXY,Yao2025LRXY,Liu2026Percolation2,Xiao2026LERW,Xiao2025Sak,Xiao2025Heisenberg}, have consistently supported $\sigma_* = 2$.

In this paper, we investigate the 2D LR XY model in the nonclassical regime $\sigma \leq 2$.
First,  the Gaussian spin-wave theory shows that the low-$T$ phase at the marginal case $\sigma=2$
is a logarithmic quasi-long-range-ordered (log-QLRO) phase, with $C(r)\sim(\ln r)^{-\eta_{\ell}}$, $\eta_{\ell}=1/(2\pi\kappa)$, and an infrared stiffness kernel $\gamma(k)\simeq \kappa k^{2}\ln(1/k)$, where $\kappa \propto 1/T$ is the reduced LR coupling. This kernel destroys true LRO and strongly suppresses vortex proliferation: the vortex-pair energy grows as $(\ln r)^{2}$, so the standard BKT unbinding mechanism is absent.
The SR algebraic QLRO phase is unstable against an arbitrarily weak LR perturbation for $\sigma\leq 2$.
In a finite $L\times L$ system, the logarithmic kernel predicts $M^{2}\sim(\ln L)^{-\eta_{\ell}}$ and $(\xi/L)^{2}\simeq \kappa \ln L$, where $M^2$ is the squared magnetization density and $\xi$ the second-moment correlation length. 
At finite temperature, vortex and anharmonic fluctuations renormalize $\kappa \to \kappa_{\rm R}$.
Nevertheless, as temperature decreases, this renormalization is suppressed and  the spin-wave predictions become asymptotically accurate.

Second,  we construct an extended BKT flow for $\sigma=2$ within the adiabatic approximation by absorbing the LR contribution into a running effective spin stiffness, $K(\ell)=K_{0}+\kappa\ell$. The flow exhibits a genuine second-order fixed point rather than the standard BKT fixed point, with $K^{*}=2/\pi$ and a finite vortex fugacity $y^{*}=\sqrt{\kappa/(16\pi)}$, yielding a nonuniversal correlation-length exponent $\nu\simeq 1/\sqrt{2\pi\kappa}$. Hence the transition is beyond both the BKT and Ginzburg-Landau-Wilson (GLW) paradigms. 

Third, large-scale Monte Carlo simulations confirm the theoretical picture. 
For \(\sigma\leq2\), an arbitrarily weak LR perturbation makes \((\xi/L)^2\) diverge with \(L\), demonstrating the instability of the SR algebraic QLRO phase.
At \(\sigma=2\), as the critical point is approached from the high-\(T\) side, the growth of the correlation length deviates from exponential BKT scaling and follows a power-law divergence, supporting a continuous transition. In the low-\(T\) log-QLRO phase, the predicted identity \(2\pi\kappa_{\mathrm R}\eta_{\ell}=1\) is quantitatively confirmed, and the logarithmic infrared kernel \(\gamma(k)\simeq\kappa_{\mathrm R}k^{2}\ln(1/k)\) is directly revealed in simulations.

These results establish the phase diagram of the 2D LR XY model: true LRO for $\sigma<2$; standard BKT algebraic QLRO for $\sigma>2$; and, at $\sigma=2$, a log-QLRO low-$T$ phase and a genuine second-order transition beyond both the BKT and GLW paradigms.
For all $\sigma>0$, there is only a single transition, with no intermediate phase.
Hence the crossover boundary between the LR and SR universality classes occurs exactly at $\sigma_{*}=2$,
rather than at $2-\eta_{\mathrm{SR}}$ as Sak's criterion would suggest, resolving the long-standing controversy over the phase diagram.

\section{Gaussian spin-wave theory of the low-temperature phase}
\label{sec:spinwave}

The low-\(T\) physics of the 2D SR XY model is governed by the Gaussian spin-wave sector. 
In the continuum limit, the reduced Hamiltonian (\(\mathcal{H} \equiv \beta H\)) is
\[
\mathcal{H} = \frac{J}{2}\int d^2 r(\nabla \theta (\mathbf{r}))^2,
\]
where \(\theta (\mathbf{r}) \in \mathbb{R}\) is the XY spin phase at position \(\mathbf{r}\), with the \(2\pi\)-periodicity ignored, and \(J\) is the reduced SR coupling strength.
We consider the 2D XY model with LR interaction, so that \(\mathcal{H} = \mathcal{H}_{\mathrm{SR}} + \mathcal{H}_{\mathrm{LR}}\), with
\[
\mathcal{H}_{\mathrm{LR}} = -\frac{1}{2}\int d^2 x d^2 y J_{\mathrm{LR}}(|\mathbf{x} - \mathbf{y}|)\cos (\theta (\mathbf{x}) - \theta (\mathbf{y})),
\]
where the LR coupling decays algebraically as
\[
J_{\mathrm{LR}}(r) = \frac{\lambda}{r^{2 + \sigma}}. 
\]

In the low-\(T\) algebraic QLRO phase of the SR system, the spin-wave analysis is nearly exact: vortex configurations are strongly suppressed by their divergent energy cost, so that the Gaussian sector provides an accurate description of the infrared physics. 
By adding the LR coupling \( J_{\mathrm{LR}} \), 
the spin stiffness is enhanced and thus the Gaussian spin-wave analysis becomes even more reliable. 

\begin{table*}
\centering
\caption{
Three regimes induced by the long-range interaction \(\sim 1/r^{2+\sigma}\): the \(\sigma\) range, the infrared stiffness kernel \(\gamma(k)\), the phase, and the spin correlation \(C(r)\). Here \(\eta\) and \(\eta_\ell\) are temperature dependent, while \(\eta_{\rm G}=2-\sigma\) holds strictly.
}
\label{tab:phases}
\begin{tabular}{llll}
\toprule
$\sigma$ range \hspace{3mm} & Infrared behavior \hspace{3mm} & Phase  \hspace{20mm} &  Correlation  \\
\midrule
$\sigma > 2$ & $\gamma(k) \sim k^2$ & Algebraic QLRO       & $C(r) \sim r^{-\eta}$ \\
$0 < \sigma < 2$ & $\gamma(k) \sim k^\sigma$ & True LRO   & $C(r) \sim C_0+a\, r^{-\eta_{\rm G}}$  \\
$\sigma = 2$ & $\gamma(k) \sim k^2 \ln(1/k)$ & Logarithmic QLRO  & $C(r) \sim (\ln r)^{-\eta_\ell}$\\
\bottomrule
\end{tabular}
\end{table*}

\subsection{Infrared stiffness kernel and log-QLRO}
Transforming to momentum space, the effective reduced Hamiltonian reads
\[
\mathcal{H} = \frac{1}{2}\int \frac{d^2k}{(2\pi)^2}\gamma (k)|\theta (\mathbf{k})|^2,
\]
where the stiffness kernel contains both SR and LR parts, \(\gamma (k) = Jk^{2} + \gamma_{\mathrm{LR}}(k)\), with
\begin{equation}
\begin{aligned}
\gamma_{\mathrm{LR}}(k) &= \lambda \int d^2 r\frac{1 - \cos(\mathbf{k}\cdot\mathbf{r})}{r^{2 + \sigma}} \\
&= 2\pi \lambda k^{\sigma} \int_{ka}^\infty \frac{d u}{u^{1 + \sigma}} [1 - J_0(u)],
\end{aligned}
\end{equation}
where \(J_{0}(x)\) is the zeroth-order Bessel function of the first kind, and for small momentum \(ka\ll 1\), $J_0(u) \simeq 1-u^2/4$. For simplicity, from now on we shall set the ultraviolet lattice cutoff \(a=1\). 
The leading  stiffness kernel is therefore
\begin{equation}
\gamma (k)\simeq \begin{cases}
B_{\sigma} k^2, & \sigma >2,\\
A_{\sigma}k^{\sigma}, & 0< \sigma < 2,\\
\kappa \, k^2\ln \left(1/k\right), & \sigma = 2,
\end{cases} \label{eq:kernel}
\end{equation}
where \(B_{\sigma}, A_{\sigma}, \kappa  = \pi \lambda /2 >0 \) are constants. 
For \(\sigma >2\), the LR contribution is subleading, and the system retains standard BKT physics with a renormalized stiffness.
For \(\sigma < 2\), the dominant LR contribution, \(\gamma_{\mathrm{LR}}(k)\sim A_{\sigma}k^{\sigma}\), is nonanalytic and leads to true LRO with Goldstone exponent \(\eta_{\mathrm{G}} = 2 - \sigma\). 

At \(\sigma = 2\), the logarithmic modification of the stiffness kernel has profound consequences. The integration of \(1 / \gamma (k)\) over momentum space yields a double-logarithmic divergence in the transverse susceptibility, indicating that long-wavelength fluctuations are strong enough to destroy spontaneous magnetization, as proved by Bruno~\cite{Bruno2001}. Meanwhile, the same logarithmic factor generates the log-QLRO phase. In this phase, the correlation function \(C(\mathbf{r}) = \langle \mathbf{S}(\mathbf{r})\cdot \mathbf{S}(0)\rangle\), with \(\mathbf{S}(\mathbf{r})\) the XY spin at position \(\mathbf{r}\), decays algebraically with the logarithm of distance as \(C(r)\sim (\ln r)^{- \eta_\ell}\), as demonstrated below.

Within the Gaussian spin-wave approximation, the correlation function is
\begin{equation}
C(\mathbf{r}) = \langle e^{i(\theta(\mathbf{r}) - \theta(0))}\rangle = 
e^{-\frac{1}{2} \langle (\theta(\mathbf{r}) - \theta(0))^2\rangle},
\label{eq:C_Gaussian}
\end{equation}
where \(\langle\cdots\rangle\) denotes the Gaussian average over the
phase field \(\theta(\mathbf{r})\) in the continuum limit.
Evaluating the integral for $\sigma = 2$, we find
\begin{align}
\langle [\theta(\mathbf{r}) - \theta(0)]^2 \rangle \simeq \frac{1}{\pi \kappa } \int_{1/r}^{\Lambda} \frac{dk}{k \ln(1/k)}  
\simeq \frac{1}{\pi \kappa } \ln \left[ \ln\left( r\right) \right], \nonumber 
\end{align}
with $\Lambda < 1$ an ultraviolet cutoff, and thus,
\begin{equation}
C(\mathbf{r}) \sim \left(\ln r\right)^{-\eta_\ell}, \qquad \eta_\ell = \frac{1}{2\pi \kappa },
\label{eq:C_log}
\end{equation}
where $\kappa = \pi \lambda/2 \propto 1/T$ and  $\eta_\ell \propto T$.
This decay is slower than any power law, and characterizes the log-QLRO phase,
which separates the true LRO phase for $\sigma < 2$ and the algebraic QLRO phase for $\sigma>2$.
The three regimes, together with the corresponding infrared kernel, phase, and spin correlation, are summarized in Table~\ref{tab:phases}.

\subsection{Vortex confinement and absence of the BKT transition}
The spin-wave analysis also determines the fate of topological defects. Consider a single vortex of unit topological charge in a finite $L \times L$ system. In the Gaussian spin-wave approximation, the LR part of the reduced vortex energy in the log-QLRO phase is
\[
E_{\mathrm{LR}} = \frac{1}{4}\int d^2 r d^2 r'J_{\mathrm{LR}}(\mathbf{r} - \mathbf{r}')\left[\theta_0(\mathbf{r}) - \theta_0(\mathbf{r}')\right]^2, 
\]
where \(\theta_0(\mathbf{r})\) is the vortex phase field. For \(\sigma = 2\), 
the total reduced vortex energy scales as
\begin{equation}
E_{v}(L)\simeq \frac{\kappa}{2} \left(\ln L\right)^{2}, 
\end{equation}
dominating the configurational entropy \(S_{v}\sim 2\ln L \). 
The reduced free energy is therefore positive and divergent:
\[
F_{v}(L)\simeq  \frac{\kappa}{2} \left(\ln L\right)^{2} - 2 \ln L \rightarrow +\infty .
\]
As a consequence, free vortices are strongly suppressed, and the standard BKT unbinding mechanism is absent. 
Similarly, for $\sigma = 2$, the LR reduced energy cost of inserting a vortex-antivortex pair with separation $r$ scales as
\begin{equation}
\Delta E_{\text{LR}}(r) \sim \left(\ln r\right)^2.
\end{equation}
This grows faster than the logarithmic potential in the SR model, leading to stronger confinement of vortex pairs.

\subsection{Stability of algebraic QLRO}
The stability of the SR algebraic QLRO phase against a weak LR perturbation can be analyzed using a controlled RG calculation within the Gaussian spin-wave theory. Because the unperturbed system is in its low-\(T\) algebraic phase, vortex excitations are irrelevant, and the RG treatment is exact within the Gaussian sector. We start from the effective action
\[
S = \frac{1}{2}\int \frac{d^2k}{(2\pi)^2}\left(Jk^2 +\kappa k^\sigma\right)|\theta (\mathbf{k})|^2, 
\]
where \(\kappa\) is the bare LR coupling. A sharp ultraviolet cutoff \(\Lambda\) is imposed, and we integrate out the fast modes in the momentum shell \(\Lambda /b< k< \Lambda\). Because the action is quadratic, this integration generates no new interactions for the slow modes. Rescaling momenta and fields to restore the cutoff,
\[
\mathbf{k}^{\prime} = b\mathbf{k},\qquad \theta^{\prime}(\mathbf{k}^{\prime}) = b^{-2}\theta (\mathbf{k}), 
\]
where the field rescaling is chosen to keep the SR stiffness \(J\) fixed, the LR term transforms as
\[
\kappa \int d^2 k k^\sigma |\theta (\mathbf{k})|^2\to \kappa b^{2 - \sigma}\int d^2 k'k'^\sigma |\theta '(\mathbf{k}')|^2. 
\]
The new LR coupling is therefore $\kappa^{\prime} = \kappa b^{2 - \sigma}$. 
Defining the logarithmic RG scale \(\ell = \ln b\), we obtain
\begin{equation}
\frac{d\kappa}{d\ell} = (2 - \sigma)\kappa .
\end{equation}
For \(\sigma < 2\), the LR interaction is relevant, driving the system to true LRO. For \(\sigma > 2\), it is irrelevant, and the system remains in the algebraic QLRO phase with an altered anomalous dimension. At \(\sigma = 2\), \(\kappa\) itself is marginal and does not flow, but the logarithmic factor gives rise to the log-QLRO phase.
These are consistent with the direct spin-wave analysis in Eq.~(\ref{eq:kernel}).

\subsection{Unusual finite-size scaling}

The log-QLRO phase at \(\sigma=2\) is characterized by the logarithmic infrared stiffness kernel \(\gamma(k)\simeq \kappa k^{2}\ln(1/k)\), which gives the spin correlation \(C(r)\sim(\ln r)^{-\eta_\ell}\) with \( \eta_\ell=1/(2\pi\kappa)\).
In a finite \(L\times L\) system, however, the finite-size scaling of the order parameter and of the lowest-momentum structure factor is more subtle than the naive replacement \(C(r)\to C(L)\) would suggest. The reason is that the physical XY spin is an exponential of the phase field, while the Gaussian action controls the phase field itself. Consequently, the spin structure factor \(\chi_L(\mathbf{k})\) is not simply proportional to the inverse stiffness kernel \(1/\gamma(k)\); it contains an additional global amplitude factor generated by the phase fluctuations.

Consider the Gaussian phase field \(\theta_{\mathbf{x}}\) on an \(L\times L\) lattice,
and denote the finite-volume stiffness kernel by \(\gamma_L(k)\), which reduces to the continuum kernel \(\gamma(k)\) for \(L \to \infty\).  
The difference at high momenta only affects nonuniversal amplitudes and subleading finite-size corrections. The finite-lattice phase propagator is then
\begin{equation}
G_L(\mathbf{r}) =\frac{1}{L^{2}}\sum_{\mathbf{q}\neq 0}
\frac{e^{i\mathbf{q}\cdot\mathbf{r}}}{\gamma_L(q)}.
\label{eq:GL}
\end{equation}
For the XY spin \(S_{\mathbf{x}}=e^{i\theta_{\mathbf{x}}}\), the spin correlation on the finite lattice is given by 
\begin{equation}
C_L(\mathbf{r})
=\langle S_{\mathbf{x}}S_{\mathbf{x}+\mathbf{r}}^{*}\rangle_L
=e^{-[G_L(0)-G_L(\mathbf{r})]},
\end{equation}
where \(\langle\cdots\rangle_L\) denotes the Gaussian average in the \(L\times L\) system. This expression is the finite-volume counterpart of the Gaussian identity used in Eq.~\eqref{eq:C_Gaussian}.
For a general allowed momentum \(\mathbf{k}\), the discrete Fourier transform is
\begin{equation}
\chi_L(\mathbf{k})
=e^{-G_L(0)}
\sum_{\mathbf{r}}e^{-i\mathbf{k}\cdot\mathbf{r}}e^{G_L(\mathbf{r})}.
\end{equation}
Expanding \(e^{G_L(\mathbf{r})}=1+G_L(\mathbf{r})+\cdots\), the constant term gives
\begin{equation}
\sum_{\mathbf{r}}e^{-i\mathbf{k}\cdot\mathbf{r}} = L^2\delta_{\mathbf{k},0},
\end{equation}
which contributes only at \(\mathbf{k}=0\). The first-order term, using Eq.~\eqref{eq:GL}, gives
\begin{equation}
\sum_{\mathbf{r}}e^{-i\mathbf{k}\cdot\mathbf{r}}G_L(\mathbf{r})
=\frac{1}{\gamma_L(k)}
\end{equation}
for \(\mathbf{k}\neq 0\), while it vanishes for \(\mathbf{k}=0\). 
The remaining terms give the higher-order convolution (h.o.c.) corrections.
On a finite lattice, the momentum sum in
\(G_L(\mathbf{r})\) is cut off at the smallest nonzero momentum
\(k_{\min}=2\pi/L\). This finite-size cutoff renders every term in the
expansion finite and eliminates the infrared divergence that would
appear in the infinite-volume limit. Consequently, the perturbative
expansion is well defined at finite \(L\), and because each additional factor of \(G_L(\mathbf{r})\) contributes a factor of \(1/[L^2\gamma_L(k_{\min})] \propto 1/\ln L\), the h.o.c.\ terms are
suppressed by additional powers of \(1/\ln L\).
Therefore, the leading behavior for the two cases is:
\begin{align}
\chi_L(\mathbf{k}) &= e^{-G_L(0)}
\left[
\frac{1}{\gamma_L(k)}+\text{h.o.c.}
\right],
\qquad \mathbf{k}\neq 0,
\label{eq:chi_k}\\[4pt]
\chi_L(0) &= e^{-G_L(0)}
\left[
L^2+\text{h.o.c.}
\right],
\qquad \;\;\;\;\;\; \mathbf{k}=0.
\label{eq:chi_0}
\end{align}
At \(\sigma=2\), the global amplitude is
\begin{equation}
e^{-G_L(0)}\sim(\ln L)^{-\eta_\ell},
\qquad
\eta_\ell=\frac{1}{2\pi\kappa},
\label{eq:amplitude}
\end{equation}
and the stiffness kernel at the lowest nonzero momentum
\(\mathbf{k}_{\min}=(2\pi/L,0)\) behaves as
\begin{equation}
\gamma_L(k_{\min})
\sim \kappa\left(\frac{2\pi}{L}\right)^{2}\ln L.
\label{eq:gamma_min}
\end{equation}

For the squared magnetization density and the Fourier counterpart,
\(M^2\equiv \chi_L(0)/L^2\) and \(M_k^2\equiv \chi_L(\mathbf{k}_{\min})/L^2\),  we have 
\begin{equation}
M^2 \sim(\ln L)^{-\eta_\ell},  \qquad
M_k^2\sim(\ln L)^{-(1+\eta_\ell)}.
\label{eq:M2Mk}
\end{equation}
Both quantities share the same global amplitude \(e^{-G_L(0)}\), and the extra logarithmic factor in \(M_k^2\) comes from the infrared stiffness kernel \(1/\gamma_L(k_{\min})\).
Moreover, from Eqs.~\eqref{eq:chi_k} and~\eqref{eq:chi_0}, 
we obtain the finite-size scaling of the squared correlation-length ratio as
\begin{equation}
\left(\frac{\xi}{L}\right)^{2}
\equiv
\frac{1}{[2L\sin(\pi/L)]^{2}}
\left(
\frac{M^2}{M_k^2}-1
\right)
\simeq \kappa \ln L,
\label{eq:xiL}
\end{equation}
where the global factor \(e^{-G_L(0)}\) is canceled and the prefactor of \(\ln L\) is given directly by the reduced LR coupling strength \(\kappa\).

The unusual character of this finite-size scaling becomes clear when
compared with the standard algebraic QLRO phase for \(\sigma>2\), 
where  the spin correlation decays as
\( C(r)\sim r^{-\eta}\),
and the finite-size scalings are
\begin{equation}
M^2\sim L^{-\eta},\qquad
M_k^2\sim L^{-\eta},\qquad
\left(\xi/L\right)^{2}\to \text{constant}. \nonumber 
\label{eq:algebraic}
\end{equation}
For the true LRO phase at \(\sigma<2\), the spin correlation approaches a constant,
\(C(r)\sim C_0+a r^{-(2-\sigma)}\), so that
\begin{equation}
M^2\to C_0,\qquad
M_k^2\sim L^{\sigma-2},\qquad
\left(\xi/L\right)^{2}\sim L^{2-\sigma}. \nonumber
\end{equation}
Thus, at the marginal case \(\sigma=2\), \((\xi/L)^2\) exhibits neither the constant behavior of algebraic QLRO nor the power-law divergence of the true LRO phase for \(\sigma<2\). It is a distinct finite-size signature of the log-QLRO phase.

\section{Extended BKT flow and critical behavior}
\label{sec:RG}

The near-exact spin-wave analysis of Sec.~\ref{sec:spinwave} establishes the low-\(T\) log-QLRO phase at \(\sigma=2\).
To describe the transition out of this phase, we now construct an RG analysis within the adiabatic approximation.
This approach treats the LR coupling as a scale-independent parameter and absorbs its effect into a running effective spin stiffness, yielding an extended set of BKT equations.

\begin{figure}[t]
\centering
\includegraphics[width=0.95\columnwidth]{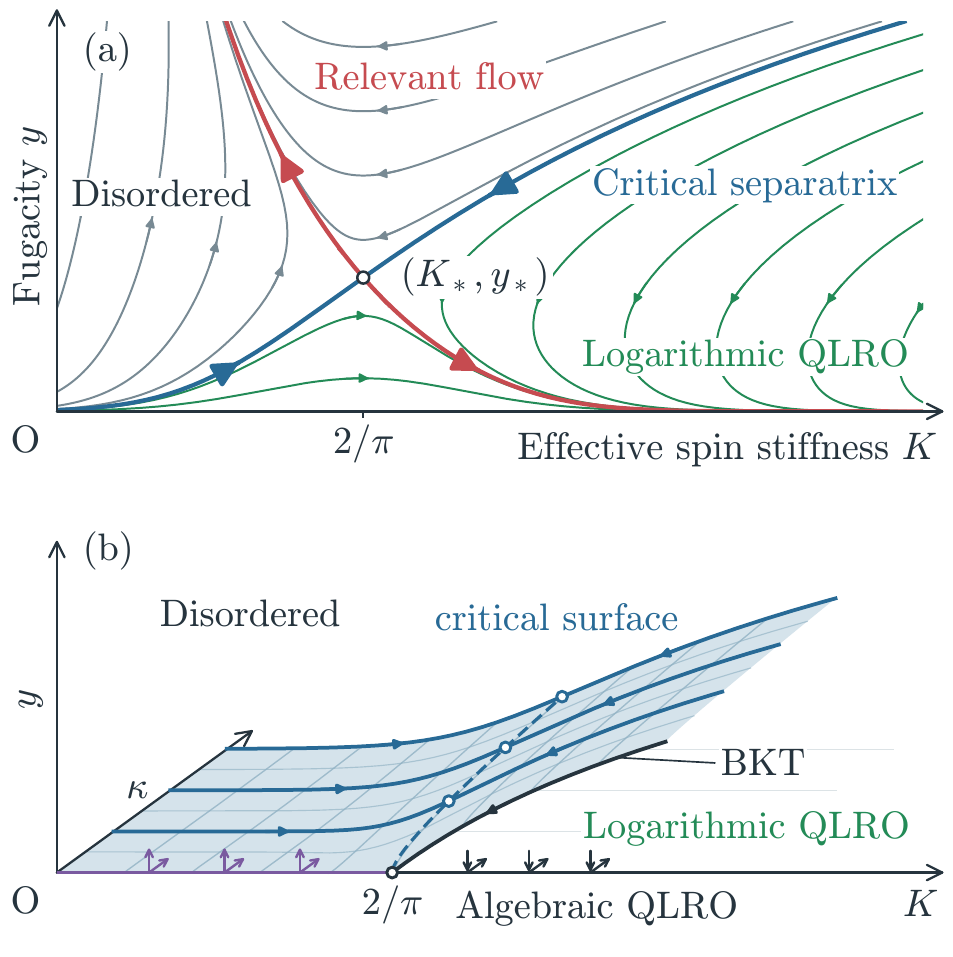}
\caption{Schematic RG flows for \(\sigma=2\), based on the extended BKT equations 
[Eq.~(\ref{eq:betay})] and the spin-wave analysis. (a) Flow in the \((K,y)\) plane for fixed \(\kappa>0\), showing the saddle fixed point, the critical separatrix (blue), and the log-QLRO and disordered phases. (b) Flow in the \((K,\kappa,y)\) space. The \(\kappa=0\) plane corresponds to the standard BKT flow; for \(\kappa>0\), the fixed points form a line whose unstable manifolds generate the critical surface (blue) separating the disordered and log-QLRO phases. In both panels, O marks the origin: \((K,y)=(0,0)\) in (a) and \((K,\kappa,y)=(0,0,0)\) in (b).
}
\label{fig:RGflows}
\end{figure}

\subsection{Extended BKT flow equations}
In the adiabatic approximation, the LR contribution to the spin-wave stiffness grows linearly with the logarithmic RG scale \(\ell = \ln(\Lambda/\mu)\), where \(\Lambda\) is the ultraviolet cutoff and \(\mu\) the running scale. We absorb this growth into a running effective stiffness
\begin{equation}
K(\ell) = K_0 + \kappa \ell ,   
\label{eq:K_running}
\end{equation}
where \(K_0\) is the bare SR stiffness and \(\kappa = \pi \lambda/2\) is the
same LR coupling introduced in Eq.~\eqref{eq:kernel}. Differentiating with respect to \(\ell\) adds a constant \(\kappa\) to the flow of \(K\), leading to the extended BKT equations
\begin{eqnarray}
\begin{aligned}
\frac{dK}{d\ell} &= \kappa -4\pi^3 K^2 y^2, \\
\frac{dy}{d\ell} &= (2 - \pi K)y, \label{eq:betay} 
\end{aligned} 
\end{eqnarray}
where \(y \) is the vortex fugacity. The structure of these equations is simple: 
for \(\kappa = 0\), they reduce to the standard BKT flow~\cite{Kosterlitz1974,Jose1977}, and the LR interaction contributes a constant \(\kappa\) to the stiffness flow,
which is scale independent because the LR coupling is marginal at \(\sigma = 2\).
Meanwhile, although the bare vortex-antivortex interaction is super-logarithmic as \(V_{\mathrm{LR}}(r)\sim (\ln r)^2\),  the effective interaction after normalization by the LR-enhanced stiffness
reduces to the standard logarithmic form \(V_{\mathrm{eff}}(r)\sim K\ln(r)\) in terms of the running stiffness \(K\). Therefore, the vortex fugacity flow retains its standard BKT form.

The two terms in the stiffness flow  \( dK/d\ell \)  have sharply different temperature dependence. 
The enhancement term is $\kappa\propto\beta$, while the softening term scales as $\sim \beta^2e^{-2c\beta}$; the latter follows from $K\propto\beta$ 
and the vortex fugacity $y\sim e^{-c\beta}$,
where $c$ is the reduced vortex-core energy. As a consequence, at low temperature,
\begin{equation}
\frac{\text{softening}}{\text{enhancement}}\sim \beta e^{-2c\beta}\to 0,
\end{equation}
so the enhancement term necessarily wins. The RG flow reinforces this conclusion: in the low-\(T\) phase $K>2/\pi$, so $dy/d\ell<0$ and $y$ decays along the flow. The softening term is therefore suppressed both by the exponentially small fugacity and by the flow itself, making the log-QLRO phase infrared stable.

At \(\sigma=2\), the LR anharmonic fluctuations are marginal at the Gaussian level,
and a more careful analysis suggests that the effective coupling scales as $T^{m-1}$ for the $m$-order term.
In other words, the one-loop contribution is a $T$-independent constant, 
and higher-loop corrections vanish as \(T\to0\) for $m \geq 2$ (see Appendix B).
Since $\kappa \propto 1/T$,
the Gaussian spin-wave predictions dominate in the low-\(T\) log-QLRO phase, with \(\kappa\) renormalized to \(\kappa_{\mathrm R}\) by vortex and anharmonic fluctuations.

\textit{Fixed point and correlation-length exponent}.
Setting the right-hand sides of Eq.~(\ref{eq:betay}) to zero gives the fixed-point conditions
\[
\kappa -4\pi^3 K^2 y^2 = 0,\qquad (2 - \pi K)y = 0. 
\]
The second equation is satisfied either by \(y = 0\) or by \(K = 2 / \pi\). If \(y = 0\), the first equation forces \(\kappa = 0\), incompatible with \(\kappa > 0\). The physically relevant fixed point therefore occurs at
\begin{equation}
 K^{*} = 2/\pi,\qquad y^{*} = \sqrt{\kappa/(16\pi)}. 
\label{eq:fixPoint}
\end{equation}
In other words, the LR interaction shifts the fixed point from the BKT line \(y = 0\) to a finite vortex fugacity, while the critical stiffness retains its universal value \(2 / \pi\).

We linearize the flow around the fixed point and define \(x = \pi K - 2\) and \(u = y - y^{*}\). Substituting into Eq.~(\ref{eq:betay}) and keeping only linear terms give
\begin{equation}
\frac{d}{d\ell}\begin{pmatrix}x\\ u\end{pmatrix} = \begin{pmatrix}-\pi\kappa & -32\pi^2 y^*\\ -y^* & 0\end{pmatrix}\begin{pmatrix}x\\ u\end{pmatrix},
\end{equation}
and the eigenvalues
\begin{equation}
\lambda_{\pm} = \frac{1}{2}\left(-\pi\kappa\pm\sqrt{\pi^2\kappa^2 + 8\pi\kappa}\right), 
\label{eq:eigenvalues}
\end{equation}
which are \(\lambda_{\pm} \simeq \pm \sqrt{2\pi\kappa}\) to leading order in small \(\kappa\). The positive eigenvalue \(\lambda_{+}\) controls the deviation of the flow from the fixed point, so the correlation length diverges as a power law, \(\xi \sim |t|^{- \nu}\), with
\begin{equation}
\nu \simeq \frac{1}{\sqrt{2\pi\kappa}}+\frac14,
\label{eq:nu-XY}
\end{equation}
where the constant $1/4$ is a subleading contribution.
The exponent is nonuniversal and depends explicitly on the LR interaction strength. As \(\kappa \rightarrow 0\), \(\nu \rightarrow \infty\), recovering the essential singularity of the BKT transition.

\textit{Anomalous dimension at criticality}.
At the fixed point, the spin-spin correlation function decays as a power law with exponent
\begin{equation}
\eta = \frac{1}{2\pi K^{*}}. 
\end{equation}
Since \(K^{*} = 2 / \pi\), one has \(\eta = 1 / 4\), unchanged from its standard BKT value. The nonzero vortex fugacity \(y^{*}\) does not affect the location of the stiffness fixed point and therefore does not alter the leading power-law exponent. Higher-order corrections from the interplay between vortices and the LR term may introduce some deviations, but within the adiabatic approximation \(\eta = 1 / 4\) holds.

\subsection{RG flow structure and critical surface}
Based on the extended BKT equations above and the low-\(T\) spin-wave analysis of Sec.~\ref{sec:spinwave}, we construct the schematic RG flows shown in Fig.~\ref{fig:RGflows}. Panel (a) shows the flow in the \((K,y)\) plane for a fixed \(\kappa > 0\). The fixed point \((K^{*},y^{*})\) is a saddle with one relevant and one irrelevant direction. The blue curve is the critical separatrix; trajectories below it flow toward \(y \rightarrow 0\) and increasing \(K\), corresponding to the log-QLRO phase, while trajectories above it flow toward growing \(y\) and decreasing \(K\), corresponding to the disordered phase. In the log-QLRO phase vortices are super-logarithmically confined, so their fugacity is irrelevantly small, whereas in the disordered phase free vortices proliferate and screen the stiffness.

Panel (b) shows the same flow in the three-parameter space \((K,\kappa ,y)\). The entire \(\kappa = 0\) plane corresponds to the SR XY model, where the RG flow reduces to the standard BKT equations, with the critical fixed point at \((K,y) = (2 / \pi ,0)\). The whole line \((\kappa = 0,y = 0)\) represents  SR spin-wave fixed points. Its stability along the \(\kappa\) direction follows directly from the spin-wave analysis at \(\sigma = 2\): any \(\kappa >0\) drives the system into the log-QLRO phase. Along the \(y\) direction, the stability is the standard BKT one: for \(K > 2 / \pi\), \(y\) flows to zero; for \(K< 2 / \pi\), \(y\) grows; and \(K = 2 / \pi\) is marginal. The union of the relevant (unstable) manifolds of the fixed-point line for \(\kappa >0\) sweeps out the critical surface shown in blue, which separates the disordered phase at large \(y\) from the log-QLRO phase at small \(y\). As \(\kappa\) increases, the critical surface rises and the log-QLRO region expands.

The critical surface for \(\kappa >0\) in Fig.~\ref{fig:RGflows}(b) corresponds to a genuine second-order phase transition, rather than the BKT transition at \(\kappa = 0\). It also transcends the GLW paradigm: below the critical surface lies the log-QLRO phase rather than true LRO; the transition is still driven by the unbinding of vortex-antivortex pairs, but in the presence of a finite vortex fugacity; and the exponent \(\nu\) is nonuniversal and depends explicitly on the LR coupling strength \(\kappa\).

\begin{figure*}[t]
\centering
\includegraphics[width=\textwidth]{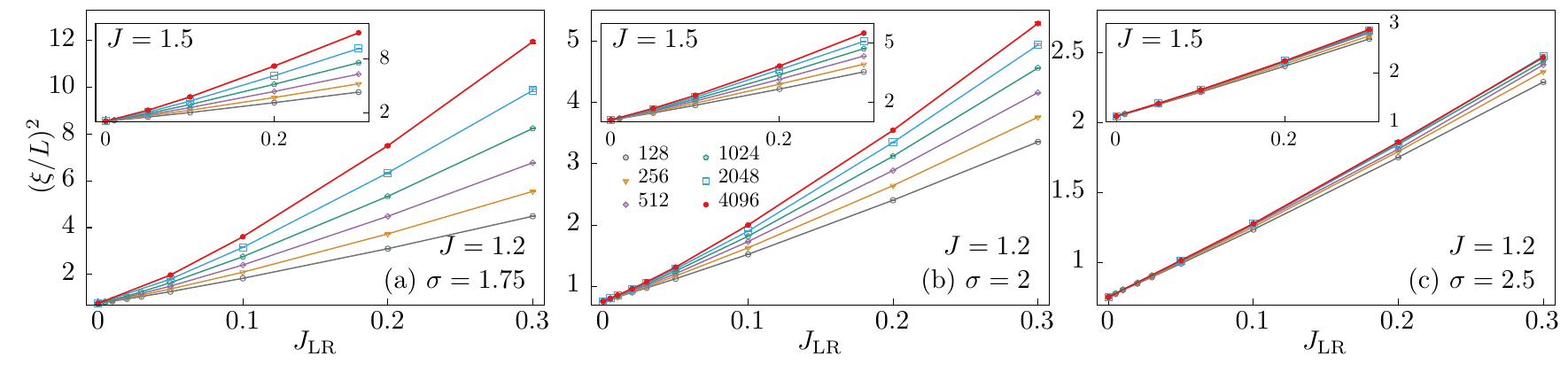}
\caption{
Instability of the SR algebraic QLRO phase against LR perturbation. The squared correlation-length ratio \((\xi/L)^2\) as a function of the LR coupling \(J_{\mathrm{LR}}\) for various system sizes \(L\) at \(J=1.2\): (a) \(\sigma=1.75\), (b) \(\sigma=2\), and (c) \(\sigma=2.5\). At \(J_{\mathrm{LR}}=0\) (open squares), the squared ratio converges to a finite constant. For \(J_{\mathrm{LR}}>0\), it diverges as \(L^{2-\sigma}=L^{0.25}\) in (a) and logarithmically as \(\ln L\) in (b), while for \(\sigma=2.5\) in (c) it converges to a finite constant, with the data for \(L=2048\) and \(L=4096\) collapsing onto a single curve. The insets are for \(J=1.5\), deep in the algebraic QLRO phase.
}
\label{fig:xi}
\end{figure*}

\section{Numerical Results}
\label{sec:num}

In this section, we present numerical evidence supporting the theoretical predictions of Secs.~\ref{sec:spinwave} and~\ref{sec:RG}. We first demonstrate the instability of the SR algebraic QLRO phase against an LR perturbation. We then show that the transition at \(\sigma=2\) is a continuous phase transition rather than a BKT transition. Finally, we directly verify the properties of the log-QLRO phase.

We consider the 2D classical XY model on an \(L\times L\) square lattice with periodic boundary conditions. The reduced Hamiltonian \(\mathcal H\equiv \beta H\) is
\begin{equation}
\mathcal H=\mathcal H_{\mathrm{SR}}+\mathcal H_{\mathrm{LR}},
\label{eq:perturbative_H}
\end{equation}
with
\begin{align}
\mathcal H_{\mathrm{SR}}&=-J\sum_{\langle i,j\rangle}\mathbf S_i\cdot\mathbf S_j,\\
\mathcal H_{\mathrm{LR}}&=-J_{\mathrm{LR}}\sum_{r_{ij}>1} \!\! f(r_{ij})\,\mathbf S_i\cdot\mathbf S_j,
\label{eq:HLR}
\end{align}
where $f(r) = b/r^{2+\sigma}$.
The SR sum runs over nearest-neighbor (NN) pairs, while the LR sum excludes NN pairs.
The constant $b$ is fixed by \(4=\sum_{r>1}f(r)\),
so the total non-NN coupling incident on a site is \(4J_{\mathrm{LR}}\). This normalization separates the strength of the LR perturbation, \(J_{\mathrm{LR}}\), from its spatial range \(\sigma\), and makes \(J_{\mathrm{LR}}/J\) the ratio of the total LR to SR coupling strengths (each site has four NN bonds).
A further advantage is that the total coupling per site is fixed for both SR and LR parts, so the energy per site of a fully polarized configuration, \(-2J-2J_{\mathrm{LR}}\), is independent of \(L\).
This reduces finite-size corrections by eliminating the \(L\)-dependent drift of the background energy density.
For \(J_{\mathrm{LR}}\to 0\), the model reduces to the standard SR XY model. We therefore consider \(J_{\mathrm{LR}}\) as a controlled perturbation to probe the stability of the SR algebraic QLRO phase for \(J > J_{\mathrm{BKT}}\), using system sizes up to \(L=4096\).

Simulations were performed using an enhanced Luijten--Bl\"ote cluster algorithm combined with the clock Monte Carlo method, which achieves \(O(1)\) computational complexity per spin update \cite{Luijten1995,Michel2019}. 
Owing to the cluster nature of the algorithm, simulations of the 2D XY model suffer little from critical slowing down; for instance, for \(L=8192\) at \(\sigma=2\) near \(T_c\), the integrated autocorrelation time is about \(9\) MC sweeps for the energy-like quantity and about \(4.6\) for the squared magnetization~\cite{Yao2025LRXY}.
For the perturbation scans of the model in Eq.~\eqref{eq:perturbative_H}, each run was thermalized for \(10^{4}\) MC sweeps. For system sizes \(L\leq 1024\), we collected more than \(10^{6}\) samples per parameter set at nonzero \(J_{\mathrm{LR}}\). For \(L=2048, 4096\), we collected more than \(10^{5}\) samples per parameter set. The shared \(J=1.2\), \(J_{\mathrm{LR}}=0\) reference at \(L=1024\) contains \(5\times 10^{5}\) samples. These counts are summed over runs. Error bars were estimated using standard binning and jackknife methods.

To characterize the phase transition and the nature of the low-\(T\) phase, we sample a broad variety of physical quantities. The magnetization
density and its Fourier mode at the smallest nonzero wave vector
\(\mathbf{k}=(2\pi/L)\hat{\mathbf{x}}\) are defined as
\begin{equation}
\mathbf{M}=\frac{1}{N}\sum_{i=1}^{N}\mathbf{S}_i,\qquad
\mathbf{M}_k=\frac{1}{N}\sum_{i=1}^{N}\mathbf{S}_i
e^{i\mathbf{k}\cdot\mathbf{r}_i},
\end{equation}
where \(N=L^2\) is the total number of spins. The squared magnetization
density is denoted by \(M^2\equiv\langle|\mathbf{M}|^2\rangle\), and,
analogously, the squared Fourier magnetization density is denoted by
\(M_k^2\equiv\langle|\mathbf{M}_k|^2\rangle\). From these, we compute the
second-moment correlation length \(\xi\) as
\begin{equation}
\xi=\frac{1}{2\sin(\pi/L)}
\sqrt{\frac{M^2}{M_k^2}-1}.
\end{equation}
The squared ratio \((\xi/L)^2\) is a powerful diagnostic of the phase type.
As the system size increases, \((\xi/L)^2\) exhibits distinct behaviors in
the four regimes: it decays as \(1/L^2\) in the disordered phase, converges
to a constant in the algebraic QLRO phase for \(\sigma>2\), and diverges in
the LRO and log-QLRO phases. The divergence takes the form \(L^{2-\sigma}\)
for \(\sigma<2\) and \(\ln L\) for \(\sigma=2\). Thus, a second-order phase
transition can be located by the asymptotically common intersection of
\((\xi/L)^2\) versus temperature for different system sizes \(L\).

In the low-\(T\) log-QLRO phase at \(\sigma=2\), the two
observables \(M^2\) and \((\xi/L)^2\) provide independent probes of the spin-wave
theory. The former determines the logarithmic decay exponent
\(\eta_\ell\) through its finite-size scaling $M^2 \sim (\ln L)^{-\eta_\ell}$, 
while the latter directly
yields the renormalized LR coupling strength
\(\kappa_{\rm R}\) via \((\xi/L)^2\simeq \kappa_{\rm R}\ln L\). 
In the Gaussian approximation, \(\kappa_{\rm R}\) reduces to the bare coupling
\(\kappa\) introduced in Sec.~\ref{sec:spinwave}; deviations from this value measure
the renormalization induced by vortex and anharmonic fluctuations. The two
determinations are tied by the spin-wave relation
\(\eta_\ell=1/(2\pi\kappa_{\rm R})\), so that measuring \(\eta_\ell\) from
\(M^2\) and \(\kappa_{\rm R}\) from \((\xi/L)^2\) provides a direct
consistency check of the Gaussian spin-wave theory in the log-QLRO
phase.

We also measure the scaled covariance ($C_{em}$ was denoted as $K$ in Ref.~\cite{Yao2025LRXY})
\begin{equation}
C_{em} = \frac{\langle E |\mathbf{M}|^2\rangle}{\langle |\mathbf{M}|^2\rangle} - \langle E\rangle,
\end{equation}
with
\begin{equation}
E = \sum_{\langle i,j\rangle}\mathbf{S}_i\cdot \mathbf{S}_j.
\end{equation}
Here \(E\) is the total sum of NN spin products and plays the role of an energy-like operator, while \(M^2\) is the square of the order parameter. 
Dimensionless universal quantities such as \(\xi/L\) are powerful and reliable for locating a continuous transition, but they are typically less efficient and suffer more severely from systematic errors for determining the correlation-length exponent \(\nu\). Rather than extracting \(\nu\) from the shapes of the \(\xi/L\) curves for different system sizes \(L\), we directly measure the linear response of the order parameter to a change in coupling via \(C_{em} = (\partial\ln\langle |\mathbf{M}|^2\rangle/\partial J)_{J_{\mathrm{LR}}}\). At a continuous transition, its finite-size scaling is $C_{em} \propto L^{y_t}$,
which provides a direct estimate of \(y_t=1/\nu\).  
For the standard BKT transition, where the correlation length diverges exponentially and no finite \(\nu\) exists, this power-law finite-size scaling is replaced by $C_{em} \propto (\ln L)^2$. 
This approach is particularly useful because it does not rely on assuming a specific form for the divergence of \(\xi\).
The \(C_{em}\) measurement therefore serves as an additional diagnostic to distinguish the continuous transition at \(\sigma=2\) from the BKT transition at \(\sigma>2\).

\subsection{Instability of the short-range algebraic QLRO phase against long-range perturbations}
\label{sec:instability}

We start with the SR XY model in the algebraic QLRO phase at reduced NN couplings \(J=1.2\) and \(1.5\), and add a small LR perturbation \(J_{\mathrm{LR}}>0\). The coupling \(J=1.2\) lies slightly above the SR BKT transition, \(J_{\mathrm{BKT}}=1.11996(6)\)~\cite{Komura2012},
and \(J=1.5\) is further inside this phase. The primary diagnostic is the squared correlation-length ratio \((\xi/L)^2\).

Figure~\ref{fig:xi} summarizes the effect of the LR perturbation.
For \(J_{\mathrm{LR}}=0\), \((\xi/L)^2\) converges to a finite constant, consistent with SR algebraic QLRO.
For any \(J_{\mathrm{LR}}>0\), it instead diverges with \(L\) for \(\sigma\le2\), signaling that the algebraic phase is destroyed.
The divergence takes the power-law form \((\xi/L)^2\sim L^{2-\sigma}=L^{0.25}\) for \(\sigma=1.75\) and the logarithmic form \((\xi/L)^2\sim \ln L \) for \(\sigma=2\), which are precisely the finite-size signatures of the LRO and log-QLRO phases, respectively, as predicted by the spin-wave theory of Sec.~\ref{sec:spinwave}.
The insets for \(J=1.5\) exhibit the same behavior.

In contrast, for \(\sigma=2.5\), \((\xi/L)^2\) converges to a finite constant for both \(J=1.2\) and \(1.5\). In particular, the data for \(L=2048\) and \(L=4096\) collapse onto a single curve, directly demonstrating algebraic QLRO and ruling out a slow divergence. Moreover, the asymptotic value of the \((\xi/L)^2\) curve depends explicitly on \(J_{\mathrm{LR}}\), directly illustrating the renormalization of the spin stiffness by the LR coupling.

Joint fits of the small-\(J_{\mathrm{LR}}\) data yield a critical LR coupling \(J_{\mathrm{LR},c}=0\) within numerical resolution, indicating that the SR fixed line is unstable against an infinitesimal LR perturbation for \(\sigma\le2\).
This result contradicts the assumption of a finite radius of convergence for the perturbative expansion in \(J_{\mathrm{LR}}\), a key assumption in Ref.~\cite{Walther2026}.

\begin{figure}[t]
\centering
\includegraphics[width=\columnwidth]{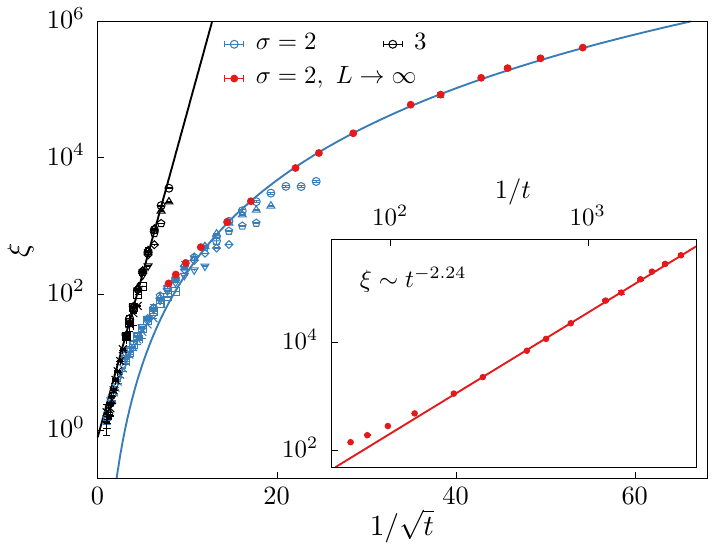}
\caption{Continuous phase transition at \(\sigma=2\) revealed by the power-law growth of \(\xi\) on the high-temperature side of \(T_c\) (data taken from Ref.~\cite{Yao2025LRXY}). 
For \(\sigma=3\), the data follow the exponential BKT scaling \(\xi\sim\exp(b/\sqrt{t})\); for \(\sigma=2\) (blue), they deviate from this form as \(T_c\) is approached. 
The red points are the extrapolated correlation lengths in the thermodynamic limit, which follow the power law \(\xi\sim t^{-\nu}\) with \(\nu\approx2.24\). 
Inset: log-log plot of \(\xi\) versus \(1/t\), showing the asymptotically straight line characteristic of this power law.}
\label{fig:continuous}
\end{figure}

\subsection{Continuous phase transition at \(\sigma=2\): Evidence against BKT}
\label{sec:continuous}

We now demonstrate that the phase transition at \(\sigma=2\) is a genuine second-order transition rather than a BKT transition. Here and in Sec.~\ref{sec:logqlro}, we consider the uniform LR XY model of Ref.~\cite{Yao2025LRXY} (studied with system sizes up to \(L=8192\)), where the interaction is
\begin{equation}
J(r)=\frac{A_L}{r^{2+\sigma}},\qquad
\sum_{\mathbf{r}\neq\mathbf{0}}J(r)=4 ,
\label{eq:uniform_J}
\end{equation}
with the sum including nearest neighbors so that temperature is measured in units of \(A_L\); in terms of Eq.~\eqref{eq:perturbative_H}, this corresponds to \(J=bJ_{\mathrm{LR}}=\beta A_L\). The critical inverse temperatures for the two representative cases discussed below are
\( \beta_c(\sigma=2)=0.7315(2)\) and 
\( \beta_c(\sigma=3)=0.899(5)\), 
as determined by finite-size scaling in Ref.~\cite{Yao2025LRXY}. These two cases provide a controlled comparison: \(\sigma=3\) lies in the SR regime where standard BKT physics is expected, while \(\sigma=2\) is the marginal case where the LR term becomes logarithmically marginal.

The first diagnostic is the growth of the second-moment correlation length \(\xi\) as $T_c$ is approached from the disordered phase. For a BKT transition, $\xi$ diverges exponentially,
\begin{equation}
\xi \sim \exp\!\left(\frac{b}{\sqrt{t}}\right),
\qquad
t=\frac{T-T_c}{T_c},
\end{equation}
with \(b\) a nonuniversal constant, whereas a second-order transition exhibits a power-law divergence
\begin{equation}
\xi \sim t^{-\nu}.
\end{equation}
Figure~\ref{fig:continuous} (data taken from Ref.~\cite{Yao2025LRXY}) presents such a comparison in a semilogarithmic plot of \(\xi\) versus \(1/\sqrt{t}\).
For \(\sigma=3\), a nearly straight line, up to the temperature where finite-size effects start to become significant, confirms the exponential BKT scaling;
it was further shown~\cite{Yao2025LRXY} that, by plotting \(a\xi\) against \(b/\sqrt{t}\) and adjusting the nonuniversal constants \(a\) and \(b\), the data for \(\sigma=2.1\), \(3\), and the NN case collapse onto a single curve.
For \(\sigma=2\) (blue), the data approximately follow the same BKT behavior only for small \(1/\sqrt{t}\), i.e., deep in the high-temperature disordered phase.
As \(T_c\) is approached, however, the deviation from the BKT line becomes increasingly pronounced.

A second, and particularly direct, piece of evidence comes from the finite-size scaling extrapolation of \(\xi\) to the thermodynamic limit. The raw simulation data for \(\xi\) are truncated at finite \(L\), and near the critical point they enter the finite-size critical window where \(\xi\sim L\). To go beyond this window, Ref.~\cite{Yao2025LRXY} used the universal step-function finite-size scaling ansatz
\begin{equation}
\frac{\xi(\beta,sL)}{\xi(\beta,\;\,L)}
= F_{\xi}\!\left(\frac{\xi(\beta,L)}{L};s\right)
+ O(\xi^{-\omega},L^{-\omega}),
\label{eq:stepFSS}
\end{equation}
where \(s\) is a fixed rescaling factor, taken as \(s=2\), and \(F_{\xi}\) is a universal function of \(\xi/L\) that depends on \(s\). The correction term \(O(\xi^{-\omega},L^{-\omega})\) is small for \(\sigma=3\), but must be retained for \(\sigma=2\) because of stronger finite-size corrections. With this ansatz, the finite-size data can be reliably extrapolated to the thermodynamic limit, reaching values as large 
as $\xi \sim 10^5$--$10^6$ (red points in Fig.~\ref{fig:continuous}).
At this scale, the asymptotic growth law can be read off directly by eye, without relying on delicate fits in the critical window. For \(\sigma=2\), the extrapolated \(\xi\) falls on a straight line in a double-logarithmic plot of \(\xi\) versus \(1/t\), giving (inset of Fig.~\ref{fig:continuous})
\begin{equation}
\nu \approx 2.24.
\end{equation}
Since this estimate is obtained directly from the plot rather than from a detailed fit, no quantitative error bar is given.
For $\xi \lesssim 3 \times 10^3$, the extrapolated values agree well with the original simulation data (Fig.~\ref{fig:continuous}), validating the extrapolation procedure.

The third piece of evidence is an independent determination of \(\nu\) from the finite-size scaling of the scaled covariance \(C_{em}\propto L^{y_t}\) with \(y_t=1/\nu\). 
For \(\sigma=2\), a least-squares fit performed in Ref.~\cite{Yao2025LRXY}, using
\(C_{em} = L^{y_t} (a + b L^{-\omega})\) with \(\omega\) fixed at \(0.5\),
gives \(y_t=0.416(6)\) and \(\nu = 2.40(4)\).
We note, however, that this error estimate may be underestimated by a factor of up to about \(2\), since it relies on fixing \(\omega=0.5\).
Nevertheless, this value is consistent with the estimate \(\nu\approx 2.24\) read off directly from the thermodynamic-limit power-law growth of \(\xi\).
The agreement between the two independent determinations—one based on the extrapolated correlation length, the other on the finite-size scaling of a linear-response quantity—makes the second-order nature of the transition at \(\sigma=2\) particularly robust.
Moreover, at $\sigma=2$, the anomalous dimension at criticality was estimated to be $\eta=0.260(5)$ in Ref.~\cite{Yao2025LRXY}, which is, within two error bars, consistent with the extended-BKT prediction $\eta=1/4$ in Sec.~\ref{sec:RG}.
On the other hand, the adiabatic result \(\nu\simeq1/\sqrt{2\pi\kappa}\) [Eq.~\eqref{eq:nu-XY}] is not expected to reproduce the numerical value $\nu \approx 2.24$ quantitatively near \(\beta_c\): vortex fluctuations are strongly enhanced there and \(\kappa\) itself flows under the RG (see Appendix~B). 

\begin{figure}[t]
\centering
\includegraphics[width=\columnwidth]{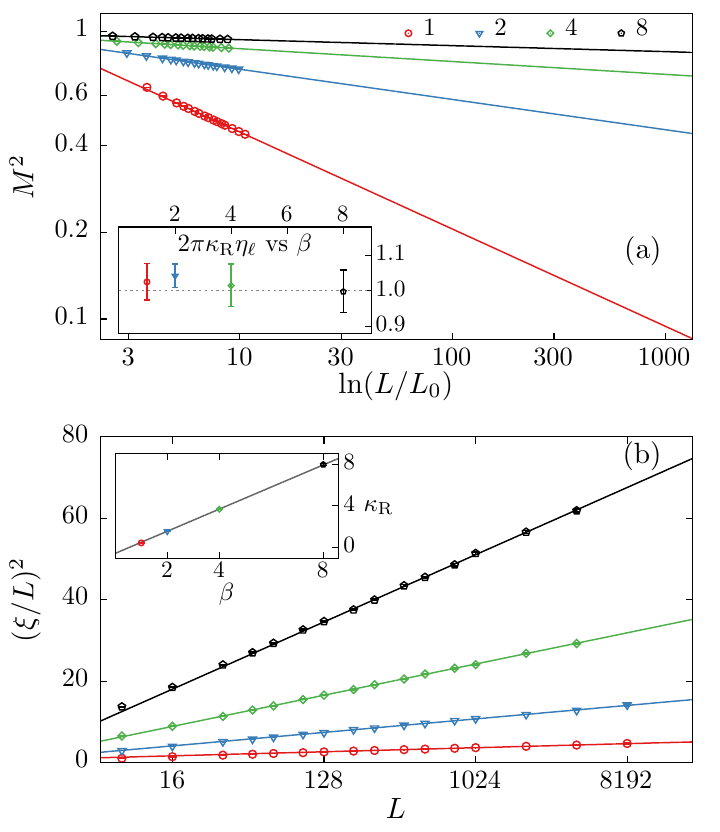}
\caption{Direct verification of the log-QLRO phase and the logarithmic stiffness kernel at \(\sigma=2\) for the uniform model. 
Panel (a): Log-log plot of $M^2$ versus \(\ln(L/L_0)\); the approximately straight lines are consistent with \( M^2 \sim(\ln L)^{-\eta_\ell}\), with \(\eta_\ell\) given by the magnitude of the slope. The inset shows the product ${\cal P} \equiv 2 \pi \kappa_{\rm R} \eta_\ell$ for $\beta=1,2,4,8$,
directly confirming the spin-wave prediction ${\cal P}=1$.
Panel (b): Semilogarithmic plot of \((\xi/L)^2\) versus \(L\); the
straight lines confirm \((\xi/L)^2\simeq \kappa_{\rm R}\ln L\), and 
the inset illustrates the linear relation of the slope \(\kappa_{\rm R} \) as a function of \(\beta\).
}
\label{fig:logqlro}
\end{figure}

\begin{table}[b]
\centering
\caption{Consistency check of the spin-wave relation
$\eta_{\ell}=1/(2\pi\kappa_{\mathrm{R}})$ in the log-QLRO phase at
$\sigma=2$, where $\eta_{\ell}$ is extracted from the finite-size scaling
of $M^{2}$; $\kappa_{\mathrm{R}}$ from the slope of $(\xi/L)^{2}$
versus $\ln L$; $\kappa$ is the bare Gaussian coupling defined in the
text. The column $1/(2\pi\kappa_{\mathrm{R}})$ gives the value of
$\eta_{\ell}$ predicted from $\kappa_{\mathrm{R}}$. The relation
$\mathcal{P}=2\pi\kappa_{\mathrm{R}}\eta_{\ell}=1$ is tested with
$\kappa_{\mathrm{R}}$, not with $\kappa$. Quoted uncertainties
include statistical fitting errors and variations under different
correction forms.}
\label{tab:logqlro}
\begin{tabular}{llllll}
\toprule
$\beta$ & $\eta_{\ell}$ & $1/(2\pi\kappa_{\mathrm{R}})$ &
$\kappa_{\mathrm{R}}$ & $\kappa$ & $\kappa_{\mathrm{R}}/\kappa$ \\
\midrule
1.0 & 0.34(1)   & 0.332(7)   & 0.48(1) & 1.043 & 0.460 \\
2.0 & 0.105(2)\; & 0.1007(13)\; & 1.58(2)\; & 2.085\; & 0.758 \\
4.0 & 0.044(2) & 0.0434(6)  & 3.67(5) & 4.170 & 0.880 \\
8.0 & 0.020(1) & 0.0200(2)  & 7.94(8) & 8.340 & 0.952 \\
\bottomrule
\end{tabular}
\end{table}

\subsection{Direct verification of the log-QLRO phase}
\label{sec:logqlro}

We now present direct numerical evidence for the log-QLRO phase and for the underlying logarithmic stiffness kernel at \(\sigma=2\).
The central spin-wave prediction is that the spin-spin correlation function decays as a power of the logarithm of distance $C(r)\sim(\ln r)^{-\eta_\ell}$.
As a result, the squared magnetization scales as
\begin{equation}
M^2\simeq  A_0 [ \ln (L/L_0)]^{-\eta_\ell}, \qquad 
\eta_\ell=\frac{1}{2\pi \kappa_{\rm R}},
\label{eq:M2log_fit}
\end{equation}
where $ \kappa_{\rm R} $ is the renormalized LR coupling strength, instead of the bare one.
A convenient way to test this prediction is to plot \( M^2 \) against \(\ln(L/L_0)\) on a log-log scale.
If Eq.~\eqref{eq:M2log_fit} holds, the data should fall on a straight line with slope \(-\eta_\ell\), 
with a magnitude growing with \(T\) (equivalently, decreasing with \(\beta\)).
This is precisely what Fig.~\ref{fig:logqlro}(a) shows.
For each temperature, we fit the data to Eq.~\eqref{eq:M2log_fit} (see Appendix~\ref{app:logqlro-fits}), with the estimated values of \(\eta_\ell\) summarized in Table~\ref{tab:logqlro}.
At sufficiently low temperatures, the results are consistent with the linear \(T\) dependence, while at intermediate temperatures, corrections become visible.

A direct probe of the infrared stiffness kernel is provided by the
squared correlation-length ratio. Figure~\ref{fig:logqlro}(b) plots
\((\xi/L)^2\) versus \(L\) on a semilogarithmic scale for several
temperatures. The data are linear in \(\ln L\), 
and the slope of each line directly yields \(\kappa_{\rm R}\), according to the
spin-wave prediction \((\xi/L)^2\simeq \kappa_{\rm R}\ln L\) of Sec.~\ref{sec:spinwave}.

The two independent determinations, $\eta_\ell$ from \(M^2\) 
and \(\kappa_{\rm R}\) from \((\xi/L)^2\), can be combined as a consistency
check of the Gaussian spin-wave theory, which predicts 
\(\eta_\ell=1/(2\pi\kappa_{\rm R})\). 
Table~\ref{tab:logqlro} summarizes the results, in which 
the two estimates of \(\eta_\ell\) agree within uncertainties.
The inset of Fig.~\ref{fig:logqlro}(a) shows that, for all $\beta=1,2,4,8$,
the product ${\cal P} \equiv 2 \pi \kappa_{\rm R} \eta_\ell$ agrees well with ${\cal P}=1$.
These results quantitatively confirm the spin-wave description of the log-QLRO phase.

Let us now compute the bare $\kappa$ directly from the LR coupling in Eq.~\eqref{eq:uniform_J}. At $\sigma=2$, in the thermodynamic limit
$A_L\to A_\infty$, $A_\infty=\frac{4}{\sum_{\mathbf{r}\neq0}r^{-4}} =\frac{6}{\pi^{2}G}\simeq0.6637$, where $G\simeq0.91597$ is the Catalan constant. The LR coefficient entering the infrared kernel is therefore $\lambda=\beta A_\infty$.
Hence the bare Gaussian coupling is
\begin{equation}
    \kappa=\frac{\pi}{2}\lambda
    =\frac{\pi}{2}\beta A_\infty
    \simeq1.0425\,\beta .
    \label{eq:kappa_bare}
\end{equation}
For $\beta=1,2,4,8$, this gives
$\kappa\simeq 1.0425$, $2.0851$, $4.1702$, $8.3403$, respectively.
This is the spin-wave kernel coefficient before vortex
renormalization. The coupling $\kappa_{\mathrm{R}}$ in
Table~\ref{tab:logqlro} is instead extracted from the slope of
$(\xi/L)^{2}$ versus $\ln L$, and therefore includes the softening
induced by vortex and anharmonic fluctuations. The identity
$\mathcal{P}=2\pi\kappa_{\mathrm{R}}\eta_{\ell}=1$ is tested with
$\kappa_{\mathrm{R}}$, not with $\kappa$. 
The values in Table~\ref{tab:logqlro} show $\kappa_{\mathrm{R}}<\kappa$, 
and further display the asymptotic $T\to 0$ tendencies that $\kappa_{\mathrm{R}}/\kappa \to 1$  
and $\kappa-\kappa_{\mathrm{R}}\approx {\rm constant}$;
this approximately constant shift can be attributed to the one-loop contribution,
as predicted in Appendix B.
Therefore, the reliability of the Gaussian spin-wave analysis in the low-\(T\) log-QLRO phase 
is directly confirmed.

\begin{figure}[t]
    \centering
    \includegraphics[width=\linewidth]{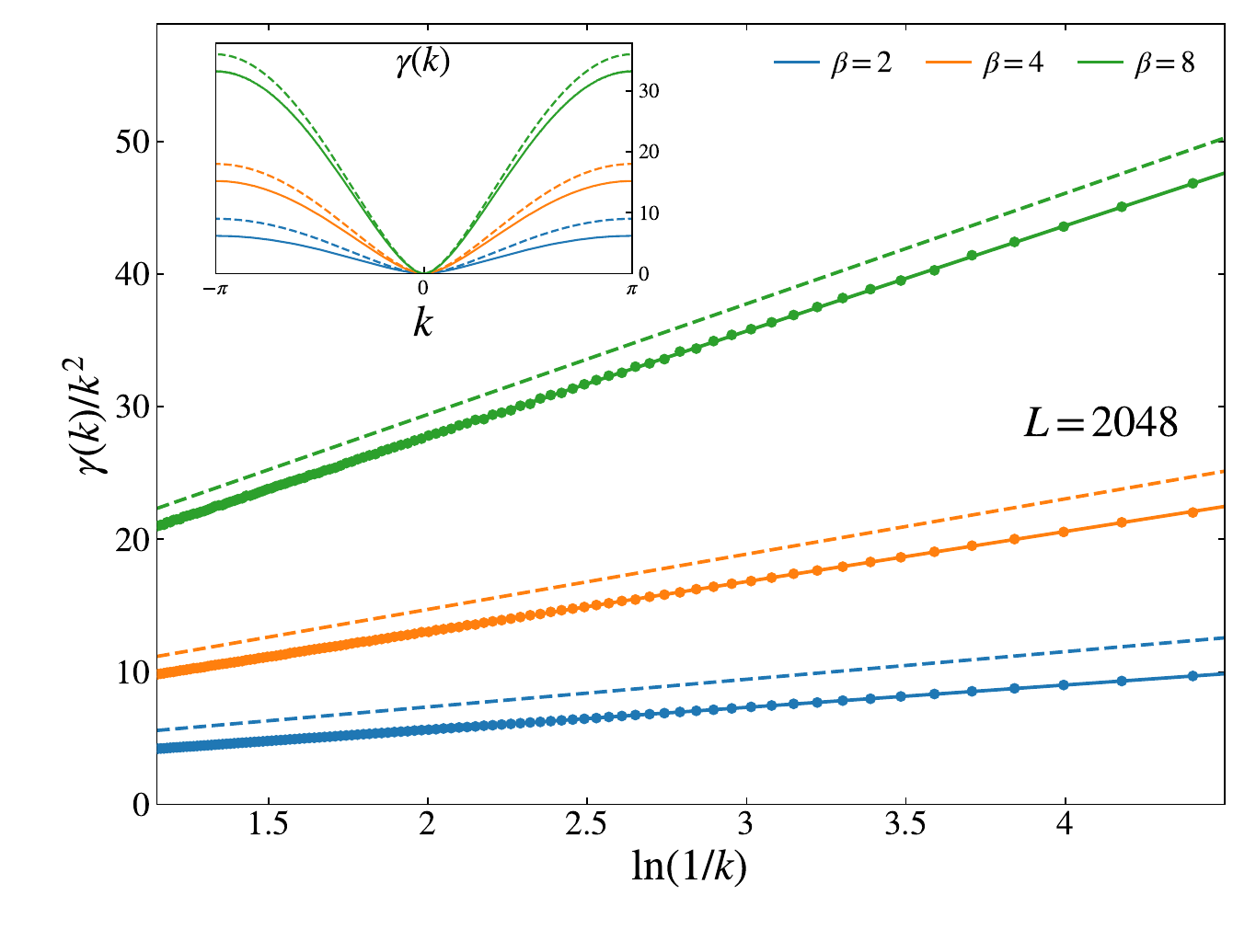}
    \vspace{-20pt}
\caption{Logarithmic infrared kernel \(\gamma(k)=M^{2}/\chi_L(k)\) of the 2D LR XY model at \(\sigma=2\). The main panel shows \(\gamma(k)/k^{2}\) against \(\ln(1/k)\) for \(L=2048\) and \(\beta=2,4,8\). Simulation data (symbols) are compared with linear fits (solid lines) and with the bare \(\kappa\) (dashed lines), computed from the LR couplings. The inset displays \(\gamma(k)\) over the full Brillouin zone \(k\in[-\pi,\pi]\).
}
    \label{fig:kernel}
    \vspace{-10pt}
\end{figure}

To probe the infrared kernel directly, we extract the effective spin-wave stiffness \(\gamma(k)=M^{2}/\chi_L(k)\) [Eqs.~\eqref{eq:chi_k} and \eqref{eq:chi_0}] from the simulation data, where \(\chi_L(k)\) is evaluated along the full momentum axis rather than at the single smallest momentum \(k_{\min}=(2\pi/L,0)\) used for \(\xi\). Figure~\ref{fig:kernel} presents \(\gamma(k)/k^{2}\) as a function of \(\ln(1/k)\) for the uniform XY model at \(L=2048\) and \(\beta=2,4,8\), with the inset showing the full dispersion of \(\gamma(k)\) across the Brillouin zone. Straight-line fits to the data confirm the logarithmic enhancement \(\gamma(k)/k^{2}\simeq \kappa_{\mathrm{R}}\ln(1/k)\). The dashed lines show the bare \(\kappa\) obtained from the LR couplings for comparison. With increasing \(\beta\), the fitted slope \(\kappa_{\mathrm{R}}\) moves toward the bare \(\kappa\), reflecting the suppression of renormalization at low temperatures and supporting the spin-wave picture.

\section{Spin-wave challenges to Sak's criterion}
\label{sec:onSak}
In 3D, the $O(N)$ spin model with SR interactions undergoes a continuous transition into an LRO phase, in which the spin-wave theory is nearly exact. With LR interactions, the stiffness kernel $\gamma(k)$ takes the form $\sim k^2$ for $\sigma>2$, $k^\sigma$ for $\sigma<2$, and $k^2\ln(1/k)$ for $\sigma=2$. The resulting spin correlation function in the LRO phase $(d>2)$ behaves as
 \[
     C(r) \simeq
    \begin{cases}
        C_0 + a\, r^{\sigma - d}, & \sigma < 2, \\
         C_0 + a \, r^{2-d} /\ln r, & \sigma = 2, \\
         C_0 + a\, r^{2-d}, & \sigma > 2,
     \end{cases}
 \]
 where $C_0>0$ represents the LRO contribution, and the subleading term arises from Goldstone-mode fluctuations. For $\sigma<2$, the algebraic correction has an exponent $\sigma-d$; for $\sigma>2$, the correction is the conventional SR form $r^{2-d}$, which in 3D is $\sim 1/r$; at $\sigma=2$, the algebraic correction acquires a logarithmic factor $1/\ln r$. Throughout the low-\(T\) ordered phase, $\sigma=2$ separates the LR and SR regimes for any $d>2$.

 At criticality, the Sak criterion~\cite{Sak1973} replaces this boundary by \(\sigma_*=2-\eta_{\mathrm{SR}}\), on the basis of the perturbative expansion around \(\sigma=d/2\)~\cite{Fisher1972}. 
 For \(d>2\), however, the spin-wave analysis shows that \(\sigma=2\) separates the LR and SR regimes throughout the low-\(T\) ordered phase. 
 If Sak's criterion were correct, then for \(\eta_{\mathrm{SR}}>0\) this would leave a window \(2-\eta_{\mathrm{SR}}<\sigma<2\) in which the ordered phase is LR-like but the critical point is SR-like. 
 Such a mismatch does not seem physically natural.

 In 2D, the spin-wave analysis at $\sigma=2$ is even more striking: rather than merely modifying subleading Goldstone corrections, the logarithmic kernel separates qualitatively distinct low-\(T\) phases---true LRO for \(\sigma<2\), a log-QLRO phase 
 for $N \geq 2$ with $C(r)\sim(\ln r)^{-\eta_\ell}$ at \(\sigma=2\), and algebraic QLRO (for \(N=2\)) or disorder (for \(N\ge 3\)) for \(\sigma>2\). 
 For the XY model at \(\sigma=2\), a finite-\(T\) BKT transition is excluded, since the vortex energy diverges as $(\ln L)^2$. The adiabatic RG analysis instead predicts a continuous transition beyond the GLW paradigm, with a nonuniversal exponent $\nu$, consistent with our simulations. Furthermore, the instability of the SR algebraic QLRO phase for all \(\sigma\le 2\) rules out an intermediate SR algebraic-QLRO phase below \(\sigma=2\).

 Finally, the universal boundary \(\sigma_*=2\) proposed by Xiao \textit{et al.}~\cite{Xiao2025Universality} follows from the observation that \(\sigma=2\) separates normal diffusion from superdiffusion in LR random walks with \(P(r)\sim 1/r^{d+\sigma}\). 
 For \(t\) steps, the characteristic displacement scales as 
 \[
 R(t)\sim
 \begin{cases}
 \sqrt{t}, & \sigma>2, \\
 \sqrt{t\ln t}, & \sigma=2, \\
 t^{1/\sigma}, & \sigma<2.
 \end{cases}
 \]
 This gives \(\sigma_*=2\) independent of dimension, symmetry, and whether the system is classical or quantum. 
 By contrast, Sak's criterion \(\sigma_*=2-\eta_{\mathrm{SR}}\) depends on dimensionality and universality class; for percolation in \(3\le d<6\), \(\eta_{\mathrm{SR}}<0\) yields \(\sigma_*>2\), 
 contradicting the standard expectation that \(\sigma>2\) belongs to the SR universality class.
 The boundary \(\sigma_*=2\) is supported by high-precision simulations~\cite{Xiao2025LRXY,Yao2025LRXY,Liu2025} and 
 by field-theoretical expansions for \(O(N)\) spin, quantum \(O(N)\), as well as percolation and Lee-Yang universality classes~\cite{Li2026Ising,Li2026Quantum,Li2026Percolation}.

 From the perspective of the \(\epsilon\)-expansion results, 
 two difficulties with Sak's criterion are apparent. First, the discontinuity it was designed to remove does not exist: \(\eta\) reduces smoothly to the SR Wilson-Fisher value at \(\sigma=2\). Second, its central assumption \(\eta=2-\sigma\) fails in the strongly fluctuating regime around \(\sigma=2\).

\section{Conclusion and Outlook}
\label{sec:conclusion}

We have investigated  the 2D classical XY model with long-range interactions \(1/r^{2+\sigma}\), combining a nearly exact Gaussian spin-wave analysis, an adiabatic renormalization-group treatment, and large-scale
Monte Carlo simulations. 
Our central results are that, at the marginal case \(\sigma=2\),
the low-\(T\) phase is a logarithmic quasi-long-range ordered
(log-QLRO) phase and that the transition into it is a novel continuous
transition beyond both the BKT and GLW paradigms. In the log-QLRO
phase, the spin correlation decays as \(C(r)\sim(\ln r)^{-\eta_\ell}\), originating
from the infrared stiffness kernel
\(\gamma(k)\simeq \kappa k^2\ln(1/k)\), which simultaneously destroys
true LRO and suppresses vortex proliferation. The numerical consistency 
for the relation \(\eta_\ell=1/(2\pi\kappa_{\rm R})\),
between \(\eta_\ell\) extracted from \(M^2\) and \(\kappa_{\rm R}\) from \((\xi/L)^2\), 
provides direct support for the spin-wave picture. 
The transition out of the log-QLRO phase is a
continuous transition with a power-law divergent correlation length,
\(\xi\sim|t|^{-\nu}\) and \(\nu\simeq 1/\sqrt{2\pi\kappa}\), rather
than the exponential BKT singularity. The exponent \(\nu\) is
nonuniversal and depends explicitly on the LR coupling strength, placing
this transition beyond the GLW paradigm.

These results resolve the long-standing controversy over the phase diagram of the 2D LR XY model.
The numerical scenario locating the crossover at \(\sigma_*=2\) captures the correct global structure~\cite{Xiao2025LRXY,Yao2025LRXY}: for \(1<\sigma<2\) the LR interaction is relevant and drives the system to true LRO via a continuous transition governed by an LR Wilson-Fisher fixed point, while for \(\sigma>2\) standard BKT physics is recovered.
However,  the nature of the \(\sigma=2\) low-\(T\) phase in that scenario was misidentified: the correlation function does not approach a constant with an inverse-logarithmic correction, but instead vanishes as a power of \(\ln r\).
The distinction is subtle because the decay is extraordinarily slow, which explains why it has been overlooked in finite-size studies.

We also clarify the status of theoretical scenarios that rely on the stability of the SR universality class against LR perturbations.
Both the scenarios proposed by Giachetti \textit{et al.}~\cite{Giachetti2021} and by Walther \textit{et al.}~\cite{Walther2026} share this premise, and Sak's criterion~\cite{Sak1973} assumes that \(\eta=2-\sigma\) holds strictly in the LR regime.
Our results invalidate this premise: at \(\sigma=2\) the algebraic QLRO description itself breaks down,
and, instead, the logarithmic QLRO emerges.
Most directly, our simulations demonstrate that the SR algebraic QLRO phase is unstable against an arbitrarily weak LR perturbation for \(\sigma\le2\).
This instability of the SR fixed point under LR perturbations might be generic to other critical systems, 
and is consistent with those \((d_c-\epsilon)\) field-theoretical expansions for LR models~\cite{Li2026Ising,Li2026Quantum, Li2026Percolation}, with $d_c$ the corresponding upper spatial dimensionality, which show that \(\eta\) smoothly reduces to the SR Wilson-Fisher value at \(\sigma=2\).
A logarithmic enhancement of spin stiffness was also proposed and observed numerically for the extraordinary boundary universality class of three-dimensional $O(N)$ models~\cite{Metlitski2022Boundary,HuDengLv2021ExtraordinaryLog}.

Several directions remain open. A sharper quantitative comparison between theory and numerics requires both higher numerical precision and further field-theoretic developments. Higher-order perturbative expansions, functional RG, and conformal field theory could pin down the nonuniversal exponent \(\nu\) and the corrections to \(\eta_\ell\). 
Experimentally, although log-QLRO is difficult to distinguish from true
LRO through magnetization measurements alone, the unusual finite-size
scaling of \(M^2\), \(M_k^2\), and \((\xi/L)^2\) provides a sensitive
probe of the logarithmic stiffness kernel.
Platforms such as ultracold atoms, Rydberg arrays, and superconducting films offer promising settings for such tests. Finally, the physics at \(\sigma=2\) should be explored in other models. The phase diagram for the long-range 2D Heisenberg model ($N=3$) has recently been established in Ref.~\cite{Xiao2025Heisenberg}, and we are currently extending this investigation to general \(O(N)\) models with $N>2$ and to the quantum XY universality class.
For models with discrete symmetry, such as the Ising model, the \(\sigma=2\) regime remains a challenging and largely unexplored territory.

\begin{acknowledgments}
We are grateful to Lode Pollet, Johannes Knolle, Luis Walther, Tian-Ning Xiao, and Jun-Kai Wang for valuable discussions.
We are particularly grateful to Johannes Knolle for his suggestion to test the stability of the short-range algebraic QLRO phase against long-range perturbations. 
We acknowledge support from the National Natural Science Foundation of China (NSFC) under Grants No. 12504265 and No. 12275263, as well as the Quantum Science and Technology National Science and Technology Major Project (Grant No. 2021ZD0301900). 
\end{acknowledgments}

\appendix

\section{Finite-size fits in the log-QLRO phase}
\label{app:logqlro-fits}

We fit the squared magnetization at $\sigma=2$ to
\begin{equation}
M^2=A_0[\ln(L/L_0)]^{-\eta_\ell},
\label{eq:app-m2}
\end{equation}
with $A_0$, $L_0$, and $\eta_\ell$ as free parameters.
Each temperature is fitted separately by weighted least squares using the measurement errors. We include all available sizes $L\ge L_{\min}$; the largest sizes are $8192$ for $\beta=1,2$ and $4096$ for $\beta=4,8$. Table~\ref{tab:app-m2} summarizes the results. Reasonable fits can be obtained at $L_{\min}=128,192$ for $\beta=1,2$ and $L_{\min}=32,48$ for $\beta=4,8$. In all cases, the exponent, amplitude, and scale remain stable as $L_{\min}$ increases. The values of $\chi^2/\mathrm{DF}$ are close to or below unity.

\begin{table}[h]
\centering
\caption{Fits of $M^2$ to Eq.~\eqref{eq:app-m2} at $\sigma=2$. All available sizes $L\ge L_{\min}$ are included. Parentheses denote statistical errors from the fit covariance. The last column lists $\chi^2/\mathrm{DF}$, with $\mathrm{DF}=N-3$ for $N$ data points.}
\label{tab:app-m2}
\setlength{\tabcolsep}{2.8pt}
\renewcommand{\arraystretch}{1.1}
\begin{tabular}{c c r@{.}l r@{.}l r@{.}l c}
\hline\hline
$\beta$ & $L_{\min}$ & \multicolumn{2}{c}{$\eta_\ell$} & \multicolumn{2}{c}{$A_0$} & \multicolumn{2}{c}{$L_0$} & $\chi^2/\mathrm{DF}$ \\
\hline
1.0 & 128 & $0$ & $338(3)$ & $0$ & $972(8)$ & $0$ & $21(1)$ & $9.8/7$ \\
 & 192 & $0$ & $338(4)$ & $0$ & $97(1)$ & $0$ & $20(2)$ & $9.7/6$ \\
\hline
2.0 & 128 & $0$ & $1051(9)$ & $0$ & $940(3)$ & $0$ & $42(3)$ & $5.0/7$ \\
 & 192 & $0$ & $105(1)$ & $0$ & $940(4)$ & $0$ & $42(4)$ & $5.0/6$ \\
\hline
4.0 & 32 & $0$ & $0446(2)$ & $0$ & $9639(6)$ & $0$ & $57(2)$ & $6.8/10$ \\
 & 48 & $0$ & $0446(3)$ & $0$ & $9639(7)$ & $0$ & $57(2)$ & $6.8/9$ \\
\hline
8.0 & 32 & $0$ & $0208(1)$ & $0$ & $9809(3)$ & $0$ & $64(2)$ & $5.5/10$ \\
 & 48 & $0$ & $0208(1)$ & $0$ & $9808(4)$ & $0$ & $65(3)$ & $5.4/9$ \\
\hline\hline
\end{tabular}
\end{table}

The squared correlation-length ratio is fitted to
\begin{equation}
(\xi/L)^2=A_\xi+\kappa_{\rm R}\ln L,
\label{eq:app-q}
\end{equation}
with $A_\xi$ and $\kappa_{\rm R}$ as free parameters.
The fits use the same weighting procedure, temperatures, and largest sizes as above. All available sizes above each $L_{\min}$ are retained. Table~\ref{tab:app-q} compares $L_{\min}=384,512$ for $\beta=1$ and $L_{\min}=128,256$ for $\beta=2,4,8$.

At $\beta=1$, the larger minimum sizes yield consistent slopes, with $\chi^2/\mathrm{DF} $ values of $ 6.7/5$ and $5.1/4$. 
For $\beta=2$, increasing $L_{\min}$ lowers $\chi^2/\mathrm{DF}$ from $11.9/8$ to $7.7/6$, while the slope changes little.
The slopes at $\beta=4$ and $ \beta=8$ are also stable across the two size ranges. 
These fits support logarithmic growth of $(\xi/L)^2$, with a slope that increases with $\beta$. 

The final values and uncertainties reported in Table~\ref{tab:logqlro} are conservative estimates that account for statistical errors, variation among the fits reported in Tables~\ref{tab:app-m2} and~\ref{tab:app-q}, and additional analyses using correction terms and alternative fitting forms.

\begin{table}[h]
\centering
\caption{Fits of $(\xi/L)^2$ to Eq.~\eqref{eq:app-q} at $\sigma=2$. All available sizes $L\ge L_{\min}$ are included. Parentheses denote statistical errors from the fit covariance. The last column lists $\chi^2/\mathrm{DF}$, with $\mathrm{DF}=N-2$ for $N$ data points.}
\label{tab:app-q}
\setlength{\tabcolsep}{2.8pt}
\renewcommand{\arraystretch}{1.1}
\begin{tabular}{c c r@{.}l r@{.}l c}
\hline\hline
$\beta$ & $L_{\min}$ & \multicolumn{2}{c}{$A_\xi$} & \multicolumn{2}{c}{$\kappa_{\rm R}$} & $\chi^2/\mathrm{DF}$ \\
\hline
1.0 & 384 & $0$ & $30(4)$ & $0$ & $479(6)$ & $6.7/5$ \\
 & 512 & $0$ & $33(4)$ & $0$ & $476(6)$ & $5.1/4$ \\
\hline
2.0 & 128 & $-0$ & $35(7)$ & $1$ & $59(1)$ & $11.9/8$ \\
 & 256 & $-0$ & $2(1)$ & $1$ & $58(2)$ & $7.7/6$ \\
\hline
4.0 & 128 & $-1$ & $4(2)$ & $3$ & $69(3)$ & $6.2/7$ \\
 & 256 & $-1$ & $2(3)$ & $3$ & $67(5)$ & $4.5/5$ \\
\hline
8.0 & 128 & $-4$ & $0(5)$ & $7$ & $94(8)$ & $5.0/7$ \\
 & 256 & $-3$ & $9(8)$ & $7$ & $9(1)$ & $3.0/5$ \\
\hline\hline
\end{tabular}
\end{table}

\section{Reliability of the Spin-Wave Analysis and  Extended BKT Equations}
\label{sec:reliability}

We examine the validity of the two main approximations used in the main text---the Gaussian spin-wave expansion and the adiabatic extended BKT equations---and argue that both are asymptotically controlled as temperature decreases, while their quantitative use, particularly in the critical region, requires separate treatment.

{\it Long-range interactions make the spin-wave expansion more controlled.}
We start from the SR 2D XY model in its low-\(T\) BKT phase, where the Gaussian spin-wave expansion is already valid: the expansion parameter is
$\eta_{\mathrm{SR}}=1/(2\pi J)\le 1/4$,
and its smallness is what makes higher-order spin-wave terms irrelevant. Adding the LR interaction enhances the effective stiffness and suppresses phase fluctuations. This reduces the expansion parameter to $\eta_{\mathrm{eff}}=1/(2\pi J_{\mathrm{eff}})<\eta_{\mathrm{SR}}$, so the spin-wave expansion becomes even more controlled as the LR interaction becomes stronger. 
This is the intuitive reason why the spin-wave analysis becomes increasingly reliable as $\sigma$ decreases. It also explains why $T_c$ increases monotonically as $\sigma$ decreases: a stronger LR coupling produces a larger $J_{\mathrm{eff}}$ at a given temperature, so a higher temperature is required to unbind vortices.

Technically, the three regimes differ in which part of the stiffness kernel dominates the infrared, and therefore in the appropriate Gaussian starting point. For $\sigma>2$, the $Jk^2$ term dominates, and one may start from the SR Gaussian fixed point; the LR term only renormalizes $J\to J_{\mathrm{eff}}$. 
For $\sigma<2$, the $A_\sigma k^\sigma$ term dominates, and the correct starting point is the LR Gaussian fixed point with $[\theta]=(2-\sigma)/2>0$; phase fluctuations are bounded and true LRO emerges. For $\sigma=2$, the logarithmic kernel $\kappa k^2\ln(1/k)$ defines a distinct Gaussian fixed point with $[\theta]=0$ and $G_L(0)-G_L(\mathbf{r})\sim(2\pi\kappa)^{-1}\ln\ln r$; the correct starting point is this logarithmic fixed point, from which log-QLRO follows. These distinctions specify \emph{which} Gaussian theory to use.

{\it Distinct temperature dependence in the RG equations.}
As shown in Sec.~III A, the enhancement term in \(dK/d\ell\) dominates at low \(T\), making the log-QLRO phase infrared stable. The transition temperature is set by \(y_0(T_c)\sim y^*(T_c)=\sqrt{\kappa/(16\pi)}\): 
since \(y_0\) is exponentially small while \(y^*\) depends only algebraically on \(T\), this yields a finite \(T_c\). 
In the critical region \(y\sim y^*\), the exponential suppression disappears, and the adiabatic approximation with fixed \(\kappa\) is no longer controlled.

{\it Higher-order spin-wave terms.}
We now consider the quartic spin-wave term generated by expanding the cosine interaction. It is useful to contrast the SR and LR cases, because their Gaussian scaling dimensions differ.
For the SR model, the quartic term is a local single integral,
\begin{equation}
S_4^{\mathrm{SR}}=g_4\int d^2r\,(\nabla\theta)^4.
\end{equation}
At the $d=2$ Gaussian fixed point, $[\theta]=0$ and $[\nabla\theta]=1$, so the operator has dimension $+2$ and the coupling has dimension $[g_4]=-2$. The SR quartic term is therefore already irrelevant at the Gaussian level. 

For $\sigma=2$, the LR quartic term is instead a nonlocal double integral,
\begin{equation}
S_4^{\mathrm{LR}}=-\frac{g_4}{24}\int d^2r\,d^2r'\,
\frac{[\theta(\mathbf{r})-\theta(\mathbf{r}')]^4}{|\mathbf{r}-\mathbf{r}'|^4},
\end{equation}
with bare vertex $g_4\propto\kappa\propto\beta$. At the LR Gaussian fixed point,  $[\theta]=0$ and the numerator is dimensionless. Under rescaling $\mathbf{r}\to b\mathbf{r}$, the double-integral measure gives $b^4$ and the $1/r^4$ kernel gives $b^{-4}$, so the operator dimension is zero: the LR quartic term is marginal at the Gaussian level. 

Nevertheless, the LR quartic term is suppressed in the low-\(T\) log-QLRO phase by two mechanisms. 
First, its momentum-space kernel is inherited from the LR two-body interaction. Expanding the cosine interaction,
\begin{align}
-J_{\mathrm{LR}}(r)&\cos(\theta(\mathbf{r})-\theta(\mathbf{0}))
=\text{const}
+\frac12 J_{\mathrm{LR}}(r) [\theta(\mathbf{r})-\theta(\mathbf{0})]^2 \nonumber\\
&\quad-\frac{1}{24}J_{\mathrm{LR}}(r) [\theta(\mathbf{r})-\theta(\mathbf{0})]^4+\cdots,
\end{align}
the quadratic term gives the logarithmic stiffness kernel $\gamma_{\mathrm{LR}}(q)=\kappa q^2\ln(1/q)$, while the quartic term carries the same interaction kernel with four phase fields. Expanding $[\theta(\mathbf{r})-\theta(\mathbf{0})]^4$ in Fourier space directly gives the quartic vertex in terms of $\gamma_{\mathrm{LR}}$,
\begin{equation}
V_4(\mathbf{k}_1,\mathbf{k}_2,\mathbf{k}_3,\mathbf{k}_4) = 2\sum_{i=1}^4 \gamma_{\mathrm{LR}}(\mathbf{k}_i) - \sum_{1\le i<j\le 4}\gamma_{\mathrm{LR}}(\mathbf{k}_i+\mathbf{k}_j),
\end{equation}
where $\sum_{i=1}^4 \mathbf{k}_i = \mathbf{0}$ by momentum conservation. Under this condition, the two sums cancel whenever any $\mathbf{k}_i\to\mathbf{0}$, and since $V_4$ is built from $\gamma_{\mathrm{LR}}(q)=\kappa q^2\ln(1/q)$, the vertex is suppressed at small nonzero momenta (e.g., $k\sim 2\pi/L\to 0$). When higher-momentum modes are integrated out in a perturbative expansion, this vertex structure is therefore expected not to generate terms more singular than $\kappa k^2\ln(1/k)$, preserving the logarithmic form of the infrared kernel and only renormalizing its prefactor $\kappa$. 
Second, the one-loop correction to the Gaussian stiffness is temperature independent at leading order. The quartic coupling scales as $g_4\propto\kappa\propto 1/T$, while the Gaussian propagator is proportional to $1/\kappa\propto T$. Contracting two fields in the quartic vertex therefore gives a correction of order $g_4/\kappa\sim O(1)$. 
Similar analyses give that contributions from higher-order anharmonic terms scale as $T^{m-1}$ 
with $m>1$, vanishing as $T \to 0$.
This produces a temperature-independent contribution to $\kappa-\kappa_{\mathrm R}$, consistent with the approximately constant shift observed in Table~\ref{tab:logqlro}.
Because the LR quartic term is only marginal, possible logarithmic corrections from its accumulation at higher order are not excluded by this argument; they are expected to be small at low temperature, consistent with the observed accuracy of the spin-wave identity \(\mathcal{P}=2\pi\kappa_{\mathrm R}\eta_{\ell}=1\), although a full proof of their irrelevance would require a functional RG treatment.

{\it Range of validity.}
The arguments above apply to the low-\(T\) log-QLRO phase.
Even in this phase, \(\kappa\) is renormalized to
\(\kappa_{\mathrm{R}}\) by vortex and anharmonic fluctuations. 
This renormalization is already visible in Table~\ref{tab:logqlro}:
\(\kappa_{\mathrm{R}}\) remains below the bare \(\kappa\) by a nearly $T$-independent offset $\kappa-\kappa_{\mathrm{R}}\approx 0.4\text{--}0.5$ for $\beta\ge 2$, while the
relative deviation $1-\kappa_{\mathrm{R}}/\kappa\propto T$ between them decreases as the temperature is lowered. The
low-\(T\) expressions therefore become increasingly accurate
away from \(T_{c}\). In the critical region, \(y\sim y^{*}\), the
vortex fugacity is no longer exponentially small and the quartic term is not suppressed, 
so \(\kappa_{\mathrm{R}}\) is reduced significantly.
There the fixed-point structure and
the exponent \(\nu\) require a functional RG treatment or
higher-order calculations that allow \(\kappa\) to flow.

\bibliography{main}
\end{document}